\documentclass[preprint,12pt]{elsarticle}

\usepackage{amssymb}
\journal{Annals of Physics}

\begin{document}

\begin{frontmatter}



\title{Concise analytic solutions to the quantum Rabi model with  two arbitrary qubits}


\author{Liwei Duan$^{1}$, Shu He$^{1}$, and Qing-Hu Chen$^{1,2,*}$}

\address{
$^{1}$ Department of Physics, Zhejiang University, Hangzhou 310027,
 China \\
$^{2}$  Collaborative Innovation Center of Advanced Microstructures, Nanjing University, Nanjing 210093, China}

\begin{abstract}
Using extended coherent states, an analytical exact study has been
carried out for the quantum Rabi model (QRM) with   two arbitrary
qubits in a very concise way. The $G$-functions with $2 \times 2$
determinants are generally derived. For the same coupling constants,
the simplest $G$-function, resembling that in the one-qubit QRM, can
be obtained. Zeros of the $G$-function yield the whole regular
spectrum. The exceptional eigenvalues, which do not belong to the
zeros of the $G$ function, are obtained in the closed form. The Dark
states in the case of the same coupling can be detected clearly in a
continued-fraction technique. The present concise solution is
conceptually clear and  practically feasible  to the general
two-qubit QRM and therefore has many applications.

\end{abstract}


\end{frontmatter}



\section{Introduction}

Quantum Rabi model (QRM) describes a two-level atom (qubit) coupled to a
cavity electromagnetic mode (an oscillator)~\cite{Rabi}, a minimalist
paradigm of matter-light interactions with applications in numerous fields
ranging from quantum optics, quantum information science to condensed matter
physics. The solutions to the QRM are however highly nontrivial. Whether an
analytical exact solution even exists is uncertain for a long time. Recently,
Braak presented an analytical exact solution~\cite{Braak} to the QRM using
the representation of bosonic creation and annihilation operators in the
Bargmann space of analytical functions~\cite{Bargmann}. A so-called $G$%
-function with a single energy variable was derived yielding exact
eigensolutions, which is not in the closed form but well defined
mathematically. Alternatively, using the method of extended coherent
states (ECS), this $G$-function was recovered in a simpler, yet
physically more transparent manner by Chen \textit{et
al.}~\cite{Chen2012}. Braak's solution has stimulated extensive
research interests in the single-qubit QRM~\cite {extension}.

As quantum information resources, such as the quantum
entanglement~\cite {Nielsen} and the quantum
discord~\cite{ollivier}, can be easily stored in two qubits, two
qubits in a common cavity have potential applications in quantum
information technology. Such a model system with two qubits now can
be constructed in several solid
devices~\cite{Sillanpaa,2qubits,Dicalo}. Recently, some analytical
studies to the QRM with two qubits have been attempted within
various approaches~\cite
{Agarwal,Chilingaryan,Rodriguez,Peng,wang2014,Zhang,Zhang1}. By
using the ECS technique, a  $G$-function for the QRM with two
equivalent qubits,
resembling the  simplest one without a determinant in the single-qubit QRM~%
\cite{Braak}, was obtained~\cite{wang2014}. While in the Bargmann
representation, the $G$-function was built as a high order determinant, such
as  $8\times 8$ determinants for QRM with two different qubits~\cite
{Chilingaryan,Rodriguez,Peng}.

Practically, the QRM with  two arbitrary qubits  is the most general one, and can be
realized in experimental device with the greatest possibility. We believe
that a simpler $G$-function is more convenient to obtain the eigensolutions,
and can also shed light on many physical processes more clearly. In this
work, employing the ECS, we demonstrate a successful derivation of a very
concise $G$-function, which is just a $2\times 2$ determinant for the QRM
with  two arbitrary qubits. Furthermore, for the same coupling, the $G$%
-function even for  two different qubits can be reduced to a  simplest one
without the use of the determinant, like that in the single-qubit QRM~\cite
{Braak}.

\section{Analytical scheme to exact solutions}

The Hamiltonian of the QRM with two qubits can be generally written as~\cite
{Chilingaryan,Rodriguez,Peng}
\begin{equation}
H={\omega }d^{\dag }d+g_1\sigma _{1x}(d^{\dag }+d)+g_2\sigma _{2x}\;(d^{\dag
}+d)+\Delta _1\sigma _{1z}+\Delta _2\sigma _{2z},  \label{Hamiltonian}
\end{equation}
where $\Delta _i(i=1,2)$ is the energy splitting of the $i$-th qubit, $%
d^{\dag }$ creates one photon in the common single-mode cavity with
frequency $\omega $, $g_i$ describes the coupling strength between the $i$-th
qubit and the cavity, $\ $ $\sigma _{ix}$ and $\sigma _{iz}~$ are the usual
Pauli matrices of the $i$-th qubit. After a rotation with respect to the $y$-axis
 by an angle $\frac \pi 2,$ the Hamiltonian in the two-qubit basis $%
\left| 1,1\right\rangle $, $\left| 1,-1\right\rangle $, $\left|
-1,1\right\rangle $, and $\left| -1,-1\right\rangle $, which are eigenstates$%
\;$of $\sigma _{1z}\otimes \sigma _{2z}$, can be written as the following
symmetric matrix (in unit of $\omega =1$)
\begin{equation}
H=\left(
\begin{array}{cccc}
d^{\dag }d+g\left( d^{\dag }+d\right) & -\Delta _2 & -\Delta _1 & 0 \\
-\Delta _2 & d^{\dag }d+g^{\prime }\left( d^{\dag }+d\right) \; & 0 &
-\Delta _1 \\
-\Delta _1 & 0 & d^{\dag }d-g^{\prime }\left( d^{\dag }+d\right) & -\Delta _2
\\
0 & -\Delta _1 & -\Delta _2 & d^{\dag }d-g\left( d^{\dag }+d\right)
\end{array}
\right) ,  \label{H_matrix}
\end{equation}
where $g=g_1+g_2$ and $g^{\prime }=g_1-g_2$.

For later use, we express the wavefunction in terms of the Fock space as
\begin{equation}
\left| d\right\rangle =\sum_{n=0}^\infty \sqrt{n!}\left\{ a_n\left[ \left|
1,1\right\rangle \pm \left( -1\right) ^n\left| -1,-1\right\rangle \right]
+b_n\left[ \left| 1,-1\right\rangle \pm \left( -1\right) ^n\left|
-1,1\right\rangle \right] \right\} |n\rangle ,  \label{Wave_d}
\end{equation}
where $+(-)$ corresponds  to  even (odd) parity, $\left| n\right\rangle \ $is
the photonic number state. The Schr\"{o}dinger equation leads to the
recurrence relation
\begin{eqnarray}
a_{m+1} &=&\frac{\left[ \Delta _2\pm \Delta _1\left( -1\right) ^m\right]
b_m-\left( m-E\right) a_m-ga_{m-1}}{g\left( m+1\right) },  \nonumber \\
b_{m+1} &=&\frac{\left[ \Delta _2\pm \Delta _1\left( -1\right) ^m\right]
a_m-\left( m-E\right) b_m-g^{\prime }b_{m-1}}{g^{\prime }\left( m+1\right) }.
\label{coeff_d}
\end{eqnarray}
Note that they cannot be reduced to a linear three-term recurrence form. The
coefficients $a_n,b_n$ can be obtained in terms of two initial values of $a_0
$ and $b_0$ recursively.

In this paper, we will first study the general case of different coupling
strengths with the same cavity, then we turn to the special equal coupling
case.

\subsection{Two-qubit QRM for $g_1\neq g_2$}

To employ the ECS approach, we first perform the following pair of
Bogoliubov transformations with finite displacements
\begin{eqnarray}
A_{+}^{\dagger } &=&d^{\dagger }+g,\;A_{-}^{\dagger }=d^{\dagger }-g;
\label{operator_A} \\
B_{+}^{\dagger } &=&d^{\dagger }+g^{\prime },\;B_{-}^{\dagger
}=d^{\dagger }-g^{\prime },  \label{operator_B}
\end{eqnarray}
by which some diagonal matrix element $\ $can be reduced to the free
particle number operators plus a constant, which is very helpful for
the further study.

The wavefunction can be expanded in the Fock space of the operator $%
A_{+}^{\dagger }$
\begin{equation}
|A_{+}\rangle _{}=\sum_{n=0}^\infty \sqrt{n!}\left[ u_n^A\left|
1,1\right\rangle +z_n^A\left| -1,-1\right\rangle +v_n^A\left|
1,-1\right\rangle +w_n^A\left| -1,1\right\rangle \right] |n\rangle _{A_{+}},
\label{wave_A+}
\end{equation}
where$\;|n\rangle _{A_{+}}\;$is the number state in the $A_{+}^{\dagger }$%
-space, and termed as the ECS previously~\cite{chen1}. The Schr\"{o}dinger
equation straightforwardly gives
\begin{eqnarray}
u_m^A &=&\frac{\Delta _2v_m^A+\Delta _1w_m^A}{m-E-g^2},  \nonumber \\
v_{m+1}^A &=&-\frac{\Delta _1z_m^A+\Delta _2u_m^A-\left( m-E+g^2-2gg^{\prime
}\right) v_m^A}{\left( m+1\right) \left( g-g^{\prime }\right) }-\frac{%
v_{m-1}^A}{m+1},  \nonumber \\
w_{m+1}^A &=&-\frac{\Delta _2z_m^A+\Delta _1u_m^A-\left( m-E+g^2+2gg^{\prime
}\right) w_m^A}{\left( g+g^{\prime }\right) \left( m+1\right) }-\frac{%
w_{m-1}^A}{\left( m+1\right) },  \nonumber \\
z_{m+1}^A &=&-\frac{\Delta _1v_m^A+\Delta _2w_m^A-\left( m-E+3g^2\right)
z_m^A}{2g\left( m+1\right) }-\frac{z_{m-1}^A}{\left( m+1\right) },
\label{Coeff_A}
\end{eqnarray}
which cannot be reduced to the linear three-term recurrence relation
either. Note that if three initial coefficients $v_0^A,w_0^A,z_0^A$ are
given, all other coefficients will be uniquely determined recursively.

Considering the conserved parity, the wavefunction can be also expressed in
series expansion in the Fock space of $A_{-}^{\dagger }$

\[
\left| A_{-}\right\rangle =\sum_{n=0}^\infty (-1)^n\sqrt{n!}\left[
z_n^A\left| 1,1\right\rangle +u_n^A\left| -1,-1\right\rangle +w_n^A\left|
1,-1\right\rangle +v_n^A\left| -1,1\right\rangle \right] |n\rangle _{A_{-}}.
\]
The wavefunction for non-degenerate state should be the same, so we
have
\[
\sum_{n=0}^\infty \sqrt{n!}u_n^A|n\rangle _{A_{+}}=r\sum_{n=0}^\infty (-1)^n%
\sqrt{n!}z_n^A|n\rangle _{A_{-}};\;\;\sum_{n=0}^\infty \sqrt{n!}%
z_n^A|n\rangle _{A_{+}}=r\sum_{n=0}^\infty (-1)^n\sqrt{n!}u_n^A|n\rangle
_{A_{-}}.
\]
Projecting onto $\langle 0|$ and with the use of $\sqrt{n!}\left\langle
0\right| |n\rangle _{A_{+}}=(-1)^n\sqrt{n!}\left\langle 0\right| |n\rangle
_{A_{-}}=e^{-g^2/2}(g)^n$, we have one linear equation
\begin{equation}
G_{\pm }^{A_{}}=\sum_{n=0}^\infty \left[ u_n^A\mp z_n^A\right] g^n=0,
\label{GA}
\end{equation}
where $+(-)$ in left-hand-side is for even (odd) parity.

We can also expand the wavefunction in the Fock space of the  second displaced operators $B_{+}^{\dagger }$
\begin{equation}
\left| B_{+}\right\rangle =\sum_{n=0}^\infty \sqrt{n!}\left[ u_n^B\left|
1,1\right\rangle +z_n^B\left| -1,-1\right\rangle +v_n^B\left|
1,-1\right\rangle +w_n^B\left| -1,1\right\rangle \right] |n\rangle _{B_{+}}.
\label{Wave_B}
\end{equation}
The Schr\"{o}dinger equation yields
\begin{eqnarray}
u_{m+1}^B &=&-\frac{\Delta _2v_m^B+\Delta _1w_m^B-\left( m-E+g^{\prime
2}-2g^{\prime }g\right) u_m^B}{\left( m+1\right) \left( g^{\prime }-g\right)
}-\frac{u_{m-1}^B}{m+1},  \nonumber \\
v_m^B &=&\frac{\Delta _1z_m^B+\Delta _2u_m^B}{\left( m-E-g^{\prime 2}\right)
},  \nonumber \\
w_{m+1}^B &=&-\frac{\Delta _2z_m^B+\Delta _1u_m^B-\left( m-E+3g^{\prime
2}\right) w_m^B}{2g^{\prime }\left( m+1\right) }-\frac{w_{m-1}^B}{\left(
m+1\right) },  \nonumber \\
z_{m+1}^B &=&-\frac{\Delta _1v_m^B+\Delta _2w_m^B-\left( m-E++g^{\prime
2}+2g^{\prime }g\right) z_n^B}{\left( m+1\right) \left( g+g^{\prime }\right)
}-\frac{z_{m-1}^B}{\left( m+1\right) }.  \label{Coeff_B}
\end{eqnarray}
All coefficients for $m>0$ can be determined from three initial parameters $%
u_0^B,w_0^B,z_0^B$ recursively. Through the similar procedure, we can obtain
the second linear equation as
\begin{equation}
G_{\pm }^B=\sum_{n=0}^\infty \left[ v_n^B\mp w_n^B\right] (g^{\prime})^n=0.  \label{GB}
\end{equation}

There seems to be  $6$ initial
coefficients in Eqs. (\ref{GA}) and (\ref{GB}). Fortunately, they can be determined by coefficients in the
series expansion Eq. (\ref{Wave_d}) in the original Fock space. For example,
the same wavefunction for non-degenerate states implies $|A_{+}\rangle
_{}\varpropto |d\rangle $. Projecting onto $_{A+}\langle 0|$ yields
\begin{equation}
v_0^A=\sum_{n=0}^\infty b_n\;\left( -g\right) ^n;\;w_0^A=\pm
\sum_{n=0}^\infty b_ng^n;z_0^A=\pm \sum_{n=0}^\infty a_ng^n,  \label{initial}
\end{equation}
where we have used $_{\ A_{+}}\langle 0|n\rangle =\sqrt{\frac 1{n!}}%
e^{-g^2/2}\left( -g\right) ^n$, and removed irrelevant constants. Then we
can obtain $u_n^A\;$and$\;z_n^A$ in Eq. (\ref{GA}) through Eq. (\ref{Coeff_A}%
) recursively. Similarly, $v_0^B,w_0^B,z_0^B$ can also be obtained by
coefficients $a_n$ and $b_n$. Through Eq. (\ref{Coeff_B}), we can get all $%
v_n^B$ and $w_n^B$ in Eq. (\ref{GB}).

Inserting these coefficients into Eqs. (\ref{GA}) and (\ref{GB}), we can
arrive at two linear equations for $a_0$ and $b_0$%
\begin{eqnarray*}
G_{11}a_0+G_{12}b_0 &=&0, \\
G_{21}a_0+G_{22}b_0 &=&0,
\end{eqnarray*}
where $G_{11}\;$and $G_{21}$ are obtained from Eqs. (\ref{GA}) and (\ref{GB}%
) $\;$by\ setting $a_0=1$ and $b_0=0$ ; $G_{12}\;$and $G_{22}\;$are obtained by\
setting $a_0=0$ and $b_0=1$. The $G$-function is then defined with the following
$2\times 2$ determinant
\begin{equation}
G_{\pm }(E)=\left|
\begin{array}{ll}
G_{11} & G_{12} \\
G_{21} & G_{22}
\end{array}
\right| =0.  \label{G-function}
\end{equation}
Nonzero coefficients require the vanishing $G$-function in the real
physics problems. So the zeros of this $G$-function will give the
energy spectrum for the present model. In addition, from the first
one in Eq. (\ref {Coeff_A}) and the second one in Eq.
(\ref{Coeff_B}), we immediately know that $E_{ex}^{(1)}=m-g^2$ and
$E_{ex}^{(2)}=m-(g^{\prime })^2$ are two sets of the exceptional
solutions, which are not the zeros of the $G$-function. It is
interesting to note from the coefficients in Eqs. (\ref{Coeff_A}),
(\ref {Coeff_B}), and (\ref{initial}) that the present $G$-function
is a well defined transcendental function. Thus analytical exact
solutions have been formally found.

\begin{figure}[tbp]
\center
\includegraphics[width=8cm]{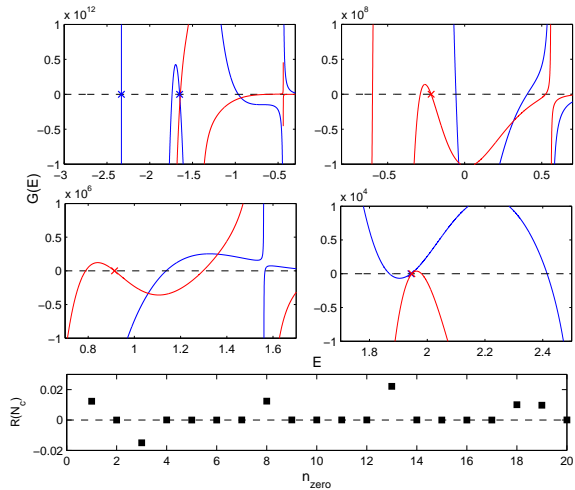}
\caption{ (Color online) (upper and middle panels) $G$-functions for the
two-qubit QRM with even (blue) and odd (red) parity using different scales
in four energy intervals connected successively. Crosses denote the unstable
zeros. (Bottom) $R_{N_c}=\ln \left( E_{N_c}/E_{N_c+1}\right) $ for different
zeros with the serial number $n_{zero}$. $\Delta _1=0.7$, $\Delta _2=0.4$
and $g_1=2g_2=0.8$. }
\label{G_function_N2_diff}
\end{figure}

We plot the $G$-function for $\Delta _1=0.7,$ $\Delta _2=0.4\;$and $%
g_1=2g_2=0.8\;$ in Fig.~\ref{G_function_N2_diff}. The stable zeros
reproduce all regular spectra, which can be confirmed by the numerical
exact solutions. Typically, the convergence is assumed to be achieved if
zeros (i.e. $E$) are determined within relative errors $\left| \left(
E_{N_c}-E_{N_c+1}\right) /E_{N_c+1}\right| <10^{-8}$, where $N_c$ is the
truncated number of the series expansions in the space of the displaced
operators. We also calculate $R_{N_c}=\ln \left( E_{N_c}/E_{N_c+1}\right) $
for all zeros, which are exhibited in Fig. \ref{G_function_N2_diff} (b). For
the stable zeros, $R_{N_c}$ is almost the same as the relative errors and
should be around the $R_{N_c}=0$ line within error $10^{-8}$, much smaller than
the symbol size.

While, very few unstable zeros, which are not the true eigenvalues, are
also present in practical calculations because of unavoidable finite
truncations. Fortunately, they can be excluded very easily. They are very
sensitive to the truncated number $N_c$ and cannot  converge with
increasing $N_c$, because the corresponding coefficients oscillate with
increasing magnitudes as $n$ increases. In sharp contrast, for the stable
zeros, the coefficients converges to zero rapidly with $n$. The positions of
unstable zeros must change even increasing $N_c$ by $1$. So they can be easily
figured out in Fig. \ref{G_function_N2_diff} (b) for apparent deviation from
$R_{N_c}=0$ line. It should be pointed out here that these unstable zeros
are absolutely not the true zeros of $G$-function, and will disappear if the
summations are really performed infinitely.

The baselines shown in Fig. \ref{G_function_N2_diff} (a) are close to $%
E_{ex}^{(1)}$ and $E_{ex}^{(2)}$ exceptional eigenvalues due to the
divergence in the $G$-functions. Generally, the exceptional eigenvalues
hardly occur for given rational model parameters.

It is interesting to note that the present $G$-function within ECS is only a
$2\times 2$ determinant, much simpler than those with $8\times 8$
determinants in the same model using the Bargmann representation~\cite{Peng}.

\subsection{Two-qubit QRM for $g_1=g_2$}

For the same coupling, $g^{\prime }=0,$ the Hamiltonian matrix (\ref
{H_matrix}) is of higher symmetry due to $H_{22}=H_{33}$, the solution will
become simpler. The wavefunction in the series expansion in the original Fock space
is the same as Eq. (\ref{Wave_d}), but the recurrence relation obtained from the
Schr\"{o}dinger equation should take a simpler form
\begin{eqnarray}
\left[ \Delta _2\pm \Delta _1\left( -1\right) ^m\right] a_m &=&\left(
m-E\right) b_m,  \label{coeff_ab} \\
\left[ \Delta _2\pm \Delta _1\left( -1\right) ^m\right] b_m &=&\left(
m-E\right) a_m+g\left[ a_{m-1}+\left( m+1\right) a_{m+1}\right] .  \nonumber
\end{eqnarray}
Note that one-to-one relation of $a_m\;$and $b_m\;$is found, and thus a
three-term recurrence relation can be obtained, in contrast with the case of
$g_1\neq g_2$. The coefficients can be obtained in terms of only one initial
value $b_0=1$ or $a_0=1$ recursively, so the continued-fraction technique is
directly applicable, similar to that in the single-qubit QRM~\cite{Swain}.
We can choose initial parameter $b_0=1$ for $\Delta _2\neq \Delta _1$,$\ $
and $a_0=1\;$for $\Delta _2=\Delta _1$ to avoid the artificial divergence.

Alternatively, we propose a mathematically well-defined technique. The
wavefunction can be expressed as Eq. (\ref{wave_A+}) in terms of the sole
pair of the displaced operators $A$ in Eq. (\ref{operator_A}). Set $%
y_m=\Delta _2v_m^A+\Delta _1w_m^A$, Eq. (\ref{Coeff_A}) can be reduced to
the following recurrence relation
\begin{eqnarray}
u_m^A &=&\frac{y_m}{  m-E-g^2  },  \nonumber \\
y_{m+1} &=&-\frac{2\Delta _2\Delta _1z_m^A+\left( \Delta _1^2+\Delta
_2^2\right) u_m^A-\left( m-E+g^2\right) y_m}{\left( m+1\right) g}-\frac{%
y_{m-1}}{m+1},  \nonumber \\
z_{m+1}^A &=&-\frac{y_m-\left( m-E+3g^2\right) z_m^A}{2g\left( m+1\right) }-%
\frac{z_{m-1}^A}{\left( m+1\right) }.  \label{r_coeff_same}
\end{eqnarray}
Although they cannot be reduced to a linear three-term relation either, the
coefficients can be uniquely given recursively by only two initial
coefficients $y_0^A$ and $z_0^A$ in this case. To this end, we can define
the $G$-function as
\begin{equation}
G_{\pm }=\sum_{n=0}^\infty \left[ u_n^A\mp z_n^A\right] g^n=0.
\label{G_fun_same}
\end{equation}

Next we need build the relationship between two sets of coefficients for
series expansions in $d$ and $A$. For the same wavefunction, we have
\[
\sum_{n=0}^\infty \sqrt{n!}y_n^A|n\rangle _{A_{+}}=r\sum_{n=0}^\infty \sqrt{%
n!}\left[ \Delta _2\pm \Delta _1\right] b_n|n\rangle ,\;\sum_{n=0}^\infty
\sqrt{n!}z_n^A|n\rangle _{A_{+}}=\pm r\sum_{n=0}^\infty \left( -1\right) ^n%
\sqrt{n!}a_n|n\rangle .
\]
Projecting onto $_{A+}\langle 0|$ yields
\begin{equation}
y_0^A=\pm \sum_{n=0}^\infty \left( \Delta _2\pm \Delta _1\right) b_n g^n,
z_0^A=\pm \sum_{n=0}^\infty a_n g^n.  \nonumber
\end{equation}
Both $y_0^A$ and $z_0^A$ are determined by coefficients $a_n$ and
$b_n$, which are only dependent on the initial parameter. Hence
$u_n^A$ and $z_n^A$ in Eq. (\ref{G_fun_same} ) can be obtained
recursively from Eq. (\ref {r_coeff_same}). Thus for the QRM of two
different qubits but with the same coupling, we have derived the
simplest  $G$-function without the use of the determinant, similar
to that in the one-qubit  QRM~\cite {Braak,Chen2012}. Note that the
$G$-function in this case in Ref.~\cite{Peng} is a $4\times 4$
determinant.

\begin{figure}[tbp]
\center
\includegraphics[width=8cm]{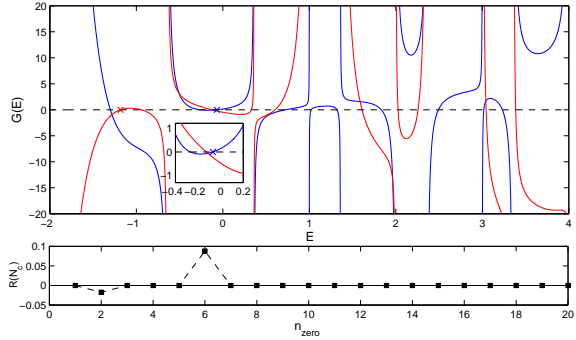}
\caption{ (Color online) (upper) $G$-functions for the two-qubit QRM with
the same coupling constants with even (blue) and odd (red) parity. Crosses
denote the unstable zeros. (Bottom) $R_{N_c}=\ln \left(
E_{N_c}/E_{N_c+1}\right) $ for different zeros with the serial number $%
n_{zero}$. $g_1=g_2=0.4$ and $\Delta _1=0.7,\Delta _2=0.4$. }
\label{G_function_N2_same}
\end{figure}

Similarly, we also plot the $G$-function
for $\Delta _1=0.7,$ $\Delta _2=0.4$ and $g_1=g_2=0.4$ in Fig.~\ref{G_function_N2_same}.  The stable zeros
give all eigenvalues and the unstable zeros can be distinguished by the same
trick outlined above.

From Eq. (\ref{r_coeff_same}), we know that $E=m-g^2$ is an
exceptional solution. However, $E=m$ is neither an exceptional
solution nor a singularity for $\ $ $\Delta _1\neq \Delta _2$, in
contrast with the previous observation~\cite{Peng}. We attribute the
difference to the possible enlarged dimension where their G function
is defined. The final results should be the same in both kinds of
treatments, but the present scheme is much more concise and allows
an in-depth discussion.

By Eq. (\ref{coeff_ab}) we may write the recurrence relation as
\begin{equation}
g\left( m+1\right) a_{m+1}=\frac{\left[ \Delta _2\pm \Delta _1\left(
-1\right) ^m\right] ^2}{\left( m-E\right) }a_m-\left( m-E\right)
a_m-ga_{m-1}.
\end{equation}
For $\Delta _2=\Delta _1=\Delta $, the analyticity of the wave
function requires $E=m$ for the even $m$ for odd parity and odd $m$
for even parity, independent of the coupling strength $g$. It is
just the trivial eigenvalue for the spin-singlet state.

For $\Delta _1\neq \Delta _2$, the first two coefficients for the states
with even parity are
\begin{equation}
a_1=-\frac 1g\left[ \left( \Delta _2+\Delta _1\right) ^2-1\right]
a_0;\;a_2=\lim_{E\rightarrow 1}\frac 1{2g}\left[ \frac{a_1}{E-1}\left(
\Delta _2-\Delta _1\right) ^2-ga_0\right] .  \nonumber
\end{equation}
The non-analyticity of the eigenfunction only occurs for the
possible divergence of $a_2$ where the denominator is zero.\ To lift the pole of $%
a_2$,  it is required that $a_1=0$, which immediately yields
\begin{equation}
\left( \Delta _2+\Delta _1\right) ^2=1.  \label{cond_1}
\end{equation}
While the first two coefficients for states with odd parity are
\begin{equation}
a_1=-\frac 1g\left[ \left( \Delta _2-\Delta _1\right) ^2-1\right]
a_0;\;a_2=\lim_{E\rightarrow 1}\frac 1{2g}\left[ \frac{a_1}{E-1}\left(
\Delta _2+\Delta _1\right) ^2-ga_0\right] .  \nonumber
\end{equation}
The analyticity of $a_2\;$requires $a_1=0,$ which gives
\begin{equation}
\left( \Delta _2-\Delta _1\right) ^2=1.  \label{cond_2}
\end{equation}
Equations (\ref{cond_1}) and (\ref{cond_2}) are just  the conditions for
special Dark states with $E=1$ found by Peng et al.,~\cite{Peng}. So they
can be easily figured out in the continued-fraction technique.

\begin{figure}[tbp]
\center
\includegraphics[width=6cm]{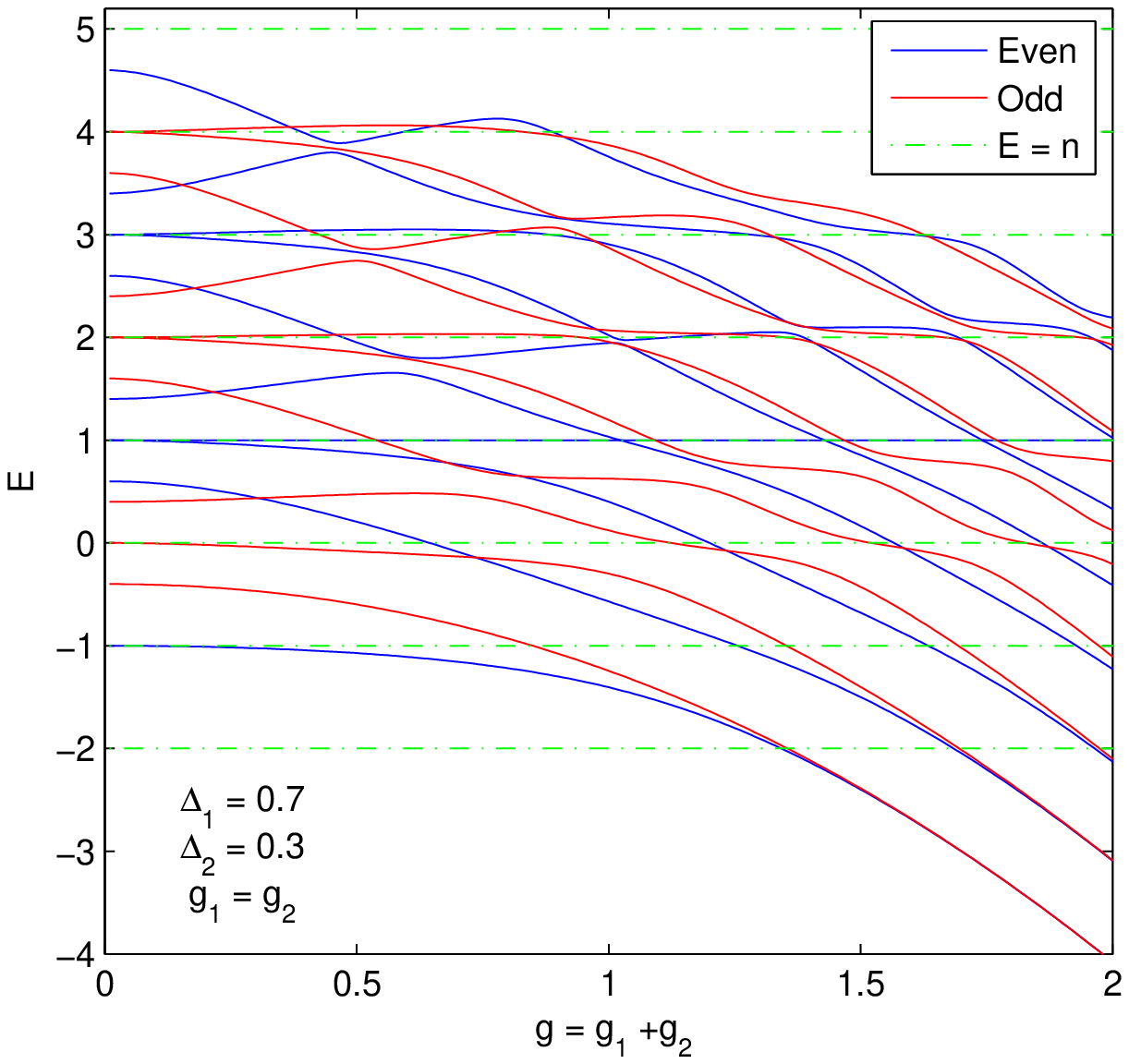} %
\includegraphics[width=6cm]{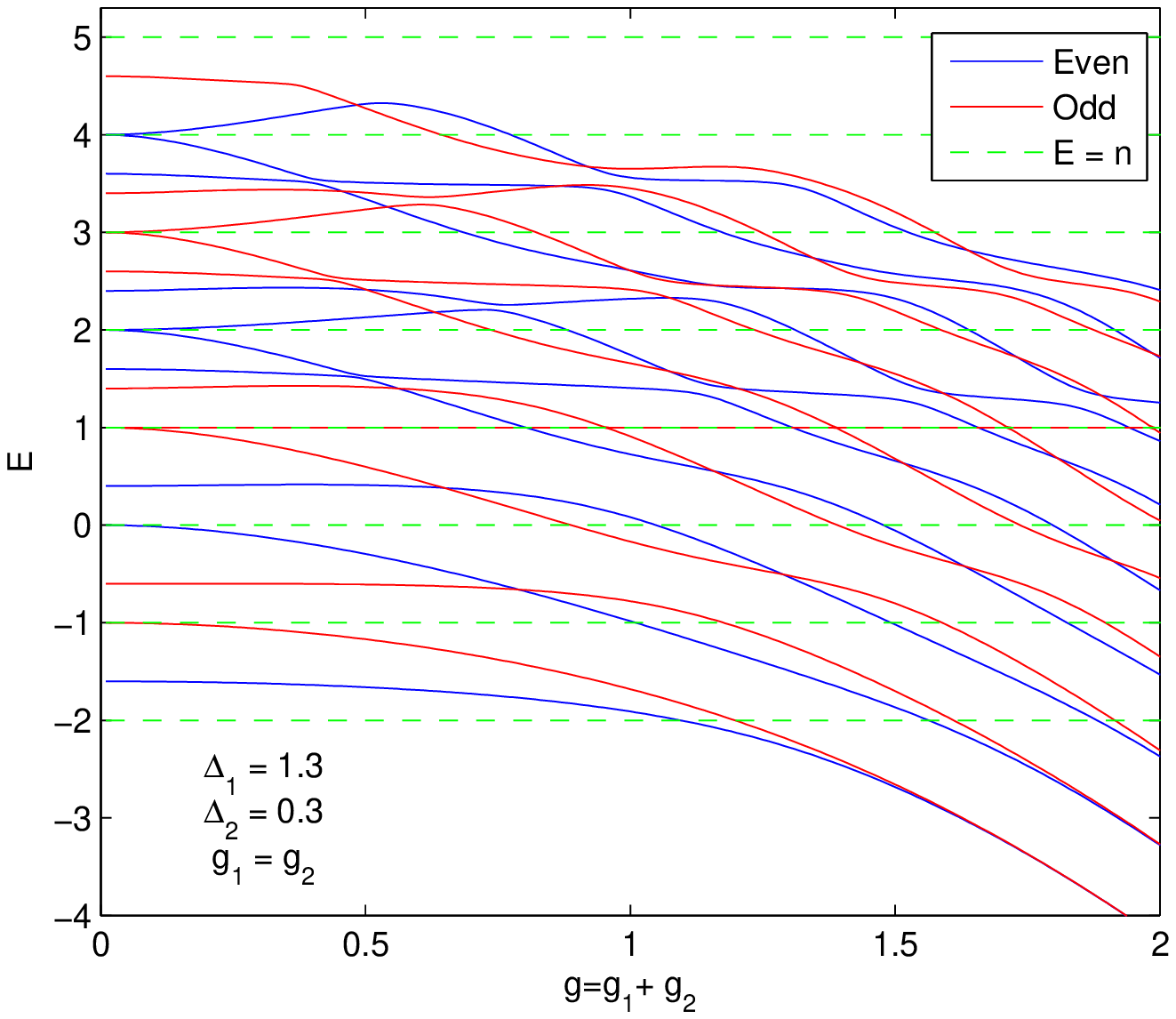}
\caption{ (Color online) The energy levels for the two-qubit QRM as a
function of the coupling $g=2g_1=2g_2$. (left) $\Delta _1=0.7,$ $\Delta
_2=0.3$; (right) $\Delta _1=1.3$ $\Delta _2=0.3$. $E=m$ denoted by dashed
lines. }
\label{energy_E1}
\end{figure}

Figure \ref{energy_E1} presents the energy spectra for two sets of
 $\Delta _1$, $\Delta _2$, which satisfy Eqs. (\ref{cond_1}) and (\ref
{cond_2}) respectively. It is shown clearly that $E=1$ for $\Delta
_1+\Delta _2=1$ and $\Delta _1-\Delta _2=1$ are $g$ independent
eigenvalues for even and odd parity respectively. The states with
$E=m\neq 1$ are not the dark states, which are  exhibited obviously
in Fig. \ref{energy_E1}.

\section{Summary and discussion}

In this work, we have derived the concise $G$-function for the QRM with
 two arbitrary qubits by using ECS, which leads to simple, analytic
solutions. Although the coefficients in the recurrence relations
look complicated, actually they can be uniquely and
straightforwardly given by the Schr\"{o}dinger equations. Our
$G$-function is only a $2\times 2 $ determinant for the general case
and a rather simple one without a determinant for the same coupling
case, much more concise than those derived recently in the Bargmann
space. This work is to extend the methodology of a compact
$G$-function in the QRM with one qubit to the QRM with  two
arbitrary qubits in the  simplest way, thereby allowing a
conceptually clear, practically feasible treatment to energy
spectra. It is our expectation that the present concise approach
will find more applications in the future.

We stress that the present analytic solution is well defined
mathematically, because of no built-in truncations, which is essentially
different from the previous finite truncation approaches~\cite
{Agarwal,Zhang,Zhang1} and the continued-fraction technique, therefore of
both fundamental and practical interest.

\textbf{ACKNOWLEDGEMENTS}

QHC acknowledge useful discussions with Daniel Braak and Jie Peng. This work
was supported by National Natural Science Foundation of China under Grant
No. 11174254, National Basic Research Program of China under Grant No.
2011CBA00103.

$^{*}$ Corresponding author. Email:qhchen@zju.edu.cn

\end{document}